\documentclass[11pt]{article}
\usepackage{amsmath,amssymb}
\usepackage[dvipdfmx]{graphicx}

\begin{document}
\begin{titlepage}
\begin{flushright}
\end{flushright}
\begin{center}
  \vspace{3cm}
  {\bf \Large  Inflationary Cosmology via Quantum Corrections in M-theory}
  \\  \vspace{2cm}
  Kazuho Hiraga and Yoshifumi Hyakutake
  \\ \vspace{1cm}
  {\it College of Science, Ibaraki University \\
   Bunkyo 2-1-1, Mito, Ibaraki 310-8512, Japan}
\end{center}

\vspace{2cm}
\begin{abstract}
We investigate inflationary cosmology by solving the effective action of the M-theory, which consists of the 11 dimensional supergravity and quartic terms of the Weyl tensor.
The metric is simply expressed by two scale factors, one for the spatial directions and the other for the internal directions.
Since the effective action of the M-theory is constructed perturbatively, we also solve the equations of motion perturbatively.
We will show that the classical solution of the 11 dimensional supergravity does not represent the inflationary expansion,
but if we include the quantum corrections, the behavior of the solution around the very early time is modified and the inflationary scenario is realized.
The inflation naturally ends when the scalar curvature becomes small and quartic terms of the Weyl tensor are negligible.
\end{abstract}

\end{titlepage}

\setlength{\baselineskip}{0.65cm}


\section{Introduction} \label{sec:Intro}


Inflationary scenario in the very early Universe provides resolution to the problems of big bang cosmology, 
such as flatness problem and horizon problem\cite{Starobinsky:1980te,Guth:1980zm,Kazanas:1980tx,Sato:1980yn}.
Numerous models have been proposed to describe the inflation, and many of them have introduced inflaton fields which slowly roll down potentials
to realize de-Sitter like vacua\cite{Linde:1983gd}-\cite{Bezrukov:2007ep} 
(see also \cite{Kolb:1990vq}-\cite{Linde:2014nna} and references there).
The origin of the inflaton field is not clear, but there are models in which the inflaton potentials are inspired 
by D-branes in superstring theory\cite{Dvali:1998pa}-\cite{McAllister:2008hb} (see also \cite{Baumann:2014nda} and references there).
Also there are a lot of models where the inflation is realized by modifying 
the gravity theory\cite{Starobinsky:1980te,Hwang:1996xh,DeFelice:2009ak,DeFelice:2010aj,Nojiri:2017ncd}.

Recently, it is possible to restrict models of the inflation by comparing cosmological parameters with the observations of Cosmic Microwave Background (CMB).
Especially observational bounds for spectral index $n_s$ and tensor to scalar ratio $r$ are measured as $n_s=0.968\pm 0.006$\cite{Ade:2015lrj}
and $r<0.07$\cite{Ade:2015tva,Array:2015xqh}.
Among other models of the inflation, the predictions of the Starobinsky model\cite{Starobinsky:1980te}, which contains curvature squared term in the action, 
are good agreement with the observations.
In fact, it predicts $n_s=0.967$ and $r=0.003$ when the number of e-folds is 60. 

Since the curvature squared term in the Starobinsky model is considered as the quantum effect of the gravity,
it is natural to ask the origin of the quantum effect in more fundamental theory, such as superstring theory or M-theory.
Actually heterotic superstring theory contains Gauss-Bonnet term, and type II superstring theories or M-theory contain quartic terms of 
the Riemann tensor\cite{Gross:1986iv}-\cite{Becker:2001pm}.
As examples, a study of the inflationary solutions in the heterotic superstring theory was done in ref.~\cite{Ishihara:1986if}, 
and studies of the inflationary solutions in the M-theory were executed in refs.~\cite{Ohta:2004wk,Maeda:2004vm,Maeda:2004hu,Akune:2006dg}.
They assumed that the scale factor for spatial directions and that for internal ones behave like $e^{H t}$ and $e^{G t}$, respectively, 
and solved equations of motion to obtain constant values of $H$ and $G$.
There are some solutions which seem to be consistent with the inflationary scenario, but in general it is difficult to obtain solutions after the inflation
which are not expressed by the ansatz of $e^{H t}$ and $e^{G t}$.

In this paper, we investigate the effective action of the M-theory which consists of the 11 dimensional supergravity and quartic terms of the Weyl tensor.
The metric is simply expressed by the scale factor $a(t)$ for the spatial directions and the scale factor $b(t)$ for the internal directions.
Since the effective action of the M-theory is constructed perturbatively, we also solve the equations of motion perturbatively.
We will show that the classical solution of the 11 dimensional supergravity does not represent the inflationary expansion,
but if we include the quantum corrections, the behavior of the solution around the very early time is modified and the inflationary scenario is realized.
The inflation naturally ends when the scalar curvature becomes small and quartic terms of the Weyl tensor are negligible.

The paper is organized as follows.
In section \ref{sec:Inf}, we review the effective action of the M-theory, and derive the equations of motion for two scale factors $a(t)$ and $b(t)$.
The 10 dimensional space directions are divided into $d$ spatial directions and $(10-d)$ internal directions.
Then we solve equations of motion perturbatively and obtain classical and quantum solutions.
In section \ref{sec:Inf4D}, we examine the case of $d=3$ in detail, and show that the inflation occurs in the early stage.
In section \ref{sec:InfanyD}, we study $d=1,\cdots,9$ in general, and confirm that the inflationary scenario is universal except for the case of $d=1$.
Section~\ref{sec:con} is devoted to conclusions and discussions.
Details of the calculations in section \ref{sec:Inf} are explicitly given in the appendix~\ref{App:cal}.


\section{Inflationary Solution} \label{sec:Inf}


\subsection{Equations of Motion with Quantum Corrections}

M-theory is a theory of membrane in 11 dimensional spacetime and its low energy limit is approximated by 11 dimensional supergravity.
There is a parameter of Planck length $\ell_p$ in the M-theory, and its effective action is expanded with respect to $\ell_p$.
From the duality between the M-theory and type IIA superstring theory, it is known that the leading correction to the supergravity starts from
$R^4$ terms, which are products of four Riemann tensors.
Actually, the bosonic part of the M-theory effective action which is relevant to the geometry 
is given by \cite{Tseytlin:2000sf,Becker:2001pm}
\begin{alignat}{3}
  S_{11} &= \frac{1}{2 \kappa_{11}^2} \int d^{11}x \; e \Big\{ R +
  \Gamma \Big(t_8 t_8 W^4 - \frac{1}{4!} \epsilon_{11} \epsilon_{11} W^4 \Big) \Big\} \notag
  \\
  &= \frac{1}{2 \kappa_{11}^2} \int d^{11}x \; e \Big\{ R + 
  24 \Gamma \big( W_{abcd} W^{abcd} W_{efgh} W^{efgh} - 64 W_{abcd} W^{aefg} W^{bcdh} W_{efgh} \notag
  \\[-0.1cm]
  &\qquad\qquad\qquad\qquad\quad
  + 2 W_{abcd} W^{abef} W^{cdgh} W_{efgh} + 16 W_{acbd} W^{aebf} W^{cgdh} W_{egfh} \label{eq:W4}
  \\
  &\qquad\qquad\qquad\qquad\quad
  - 16 W_{abcd} W^{aefg} W^b{}_{ef}{}^h W^{cd}{}_{gh} - 16 W_{abcd} W^{aefg} W^b{}_{fe}{}^h W^{cd}{}_{gh} \big) \Big\}, \notag
\end{alignat}
where $a,b,c,\cdots = 0,1,\cdots,10$ are local Lorentz indices.
Indices are lowered or raised by the flat metric $\eta_{ab}$ or $\eta^{ab}$.
$t_8$ are the products of four Kronecker's deltas with eight indices and $\epsilon_{11}$ is an antisymmetric tensor with eleven indices. 
Of course, there is a field redefinition ambiguity, so the coefficients of terms which include Ricci or scalar curvatures cannot be fixed.
In this paper, we simply deal with the effective action by employing Weyl tensor prescription.
It is confirmed that the $W^4$ terms are surely invariant under the local supersymmetry 
if we add appropriate fermionic terms\cite{deRoo:1992sm}-\cite{Hyakutake:2007sm}.

In the above effective action, there are two parameters, the gravitational constant $\kappa_{11}$ and the expansion parameter $\Gamma$,
which can be expressed as
\begin{alignat}{3}
  2\kappa_{11}^2 = (2\pi)^8 \ell_p^9, \qquad \Gamma = \frac{\pi^2\ell_p^6}{2^{11} 3^2}. \label{eq:gamma} 
\end{alignat}
The Planck length is written by $\ell_p = g_s^{1/3} \ell_s$ in terms of string length $\ell_s$ and string coupling constant $g_s$.
The numerical factor of $\Gamma$ is determined by employing the result of the 1-loop four-graviton amplitude in the type IIA superstring theory.
Since the M-theory effective action contains infinite series of higher order terms with respect to $\ell_p$,
the leading correction (\ref{eq:W4}) is reliable when $\Gamma$ is small, or in other words, 
typical length scale of a system is large compared to $\ell_p$.

Below we consider the equations of motion for the effective action (\ref{eq:W4}) up to the linear order of $\Gamma$.
By varying the effective action (\ref{eq:W4}), equations of motion are obtained as\cite{Hyakutake:2013vwa}
\begin{alignat}{3}
  E_{ab} &\equiv R_{ab} - \frac{1}{2} \eta_{ab} R 
  + \Gamma \Big\{ - \frac{1}{2} \eta_{ab} Z + R_{cdea} Y^{cde}{}_b - 2 D_{(c} D_{d)} Y^c{}_{ab}{}^d \Big\} = 0. \label{eq:MEOM}
\end{alignat}
Here $D_a$ is a covariant derivative for local Lorentz indices, and 
$X_{abcd}$, $Y_{abcd}$ and $Z$ are defined as follows.
\begin{alignat}{3}
  &X_{abcd} \!=\! 96 \big( W_{abcd} W_{efgh} W^{efgh} \!-\! 4 W_{abce} W_{dfgh} W^{efgh} \!+\! 4 W_{abde} W_{cfgh} W^{efgh} \notag
  \\
  &\quad
  \!-\! 4 W_{cdae} W_{bfgh} W^{efgh} \!+\! 4 W_{cdbe} W_{afgh} W^{efgh} \!+\! 2 W_{abef} W_{cdgh} W^{efgh} \!+\! 4 W_{ab}{}^{ef} W_{ce}{}^{gh} W_{dfgh} \notag
  \\
  &\quad
  \!+\! 4 W_{cd}{}^{ef} W_{ae}{}^{gh} W_{bfgh} \!+\! 8 W_{aecg} W_{bfdh} W^{efgh} \!-\! 8 W_{becg} W_{afdh} W^{efgh} \!-\! 8 W_{abeg} W_{cf}{}^e{}_h W_d{}^{fgh} \notag
  \\
  &\quad
  \!-\! 8 W_{cdeg} W_{af}{}^e{}_h W_b{}^{fgh} \!-\! 4 W^{ef}{}_{ag} W_{efch} W^g{}_b{}^h{}_d \!+\! 4 W^{ef}{}_{ag} W_{efdh} W^g{}_b{}^h{}_c 
  \!+\! 4 W^{ef}{}_{bg} W_{efch} W^g{}_a{}^h{}_d \notag
  \\
  &\quad
  \!-\! 4 W^{ef}{}_{bg} W_{efdh} W^g{}_a{}^h{}_c \big), \label{eq:Xdef}
  \\[0.1cm]
  &Y_{abcd} \!=\! X_{abcd} - \frac{1}{9} ( \eta_{ac} X_{bd} \!-\! \eta_{bc} X_{ad} \!-\! \eta_{ad} X_{bc} \!+\! \eta_{bd} X_{ac}) 
  + \frac{1}{90} ( \eta_{ac} \eta_{bd} \!-\! \eta_{ad} \eta_{bc} ) X, \label{eq:Ydef}
  \\[0.1cm]
  &\!Z \!=\! 24 \big( W_{abcd} W^{abcd} W_{efgh} W^{efgh}
  \!\!-\! 64 W_{abcd} W^{aefg} W^{bcdh} W_{efgh} 
  \!+\! 2 W_{abcd} W^{abef} W^{cdgh} W_{efgh} \label{eq:Zdef}
  \\
  &\quad
  \!+\! 16 W_{acbd} W^{aebf} W^{cgdh} W_{egfh}
  \!-\! 16 W_{abcd} W^{aefg} W^b{}_{ef}{}^h W^{cd}{}_{gh} 
  \!-\! 16 W_{abcd} W^{aefg} W^b{}_{fe}{}^h W^{cd}{}_{gh} \big), \notag
\end{alignat}
where $X_{ab} = X^c{}_{acb}$ and $X = X^a{}_a$.
Note that $X_{abcd} = - X_{bacd} = - X_{abdc} = X_{cdab}$ and $X_{ab} = X_{ba}$. 
It is possible to show that $W_{cdea} X^{cde}{}_b = W_{cdeb} X^{cde}{}_a$ and 
$R_{cdea}Y^{cde}{}_b = R_{cdeb}Y^{cde}{}_a$ after some calculations, and hence $E_{ab} = E_{ba}$.

Since we are interested in solutions which would inflate some of spatial directions,
we decompose 10 spatial directions into $d$ spatial ones and $(10-d)$ spatial ones. 
Then the ansatz for the metric is written as
\begin{alignat}{3}
  ds^2 &= - dt^2 + a(t)^2 dx_i^2 + b(t)^2 dy_m^2, \label{eq:metricgen}
\end{alignat}
where $i=1,\cdots,d$ and $m=d+1,\cdots,10$. 
Without loss of generality, we choose $d = 1,2,3,4$ or $5$ in the rest of this paper.
$a(t)$ and $b(t)$ are scale factors for $d$ spatial directions and $(10-d)$ spatial ones, respectively.
In this paper we regard shrinking directions as toroidal internal space, although it is interesting to compactify the internal space with nontrivial manifolds.
Substituting the ansatz (\ref{eq:metricgen}) into the eq.~(\ref{eq:MEOM}), 
we obtain differential equations for $H(t) = \dot{a}(t)/a(t)$ and $G(t) = \dot{b}(t)/b(t)$ as follows.
\begin{alignat}{3}
  E_{00} &= \frac{d(d\!-\!1)}{2} H^2 + \frac{(10\!-\!d)(9\!-\!d)}{2} G^2 + d(10\!-\!d) HG \notag
  \\
  &\quad\,
  + \Gamma \Big[ \frac{1}{2} Z - 2 d H \dot{Y}_1 - 2 (10\!-\!d) G \dot{Y}_2 
  + 2 d \big\{ \dot{H} - (d\!-\!2) H^2 - (10\!-\!d) HG \big\} Y_1 \notag
  \\
  &\quad\,
  + 2 (10\!-\!d) \big\{ \dot{G} - (8\!-\!d) G^2 - d HG \big\} Y_2 - 2 d(d\!-\!1) H^2 Y_3 - 4 d(10\!-\!d) HG Y_4 \notag
  \\
  &\quad\,
  - 2 (10\!-\!d)(9\!-\!d) G^2 Y_5 \Big], \notag
  \\
  E_{ii} &= - (d\!-\!1) \dot{H} - (10\!-\!d) \dot{G} - \frac{d(d\!-\!1)}{2} H^2 
  - \frac{(10\!-\!d)(11\!-\!d)}{2} G^2 - (d\!-\!1)(10\!-\!d) HG \notag
  \\
  &\quad\,
  + \Gamma \Big[ - \frac{1}{2} Z - 2 (\dot{H} + H^2) Y_1 + 2 (d\!-\!1) H^2 Y_3 + 2 (10\!-\!d) HG Y_4 \label{eq:MEOMgen} 
  \\
  &\quad\,
  -2 \Big\{ \frac{d}{dt} + (d\!-\!1) H + (10\!-\!d) G \Big\} \{ - \dot{Y}_1 - (d\!-\!1) H (Y_1 + Y_3) - (10\!-\!d) G (Y_1 + Y_4) \} \Big], \notag
  \\
  E_{mm} &= - d \dot{H} - (9\!-\!d) \dot{G} - \frac{d(d\!+\!1)}{2} H^2 
  - \frac{(10\!-\!d)(9\!-\!d)}{2} G^2 - d(9\!-\!d) HG \notag
  \\
  &\quad\,
  + \Gamma \Big[ - \frac{1}{2} Z - 2 (\dot{G} + G^2) Y_2 + 2 d HG Y_4 + 2 (9\!-\!d) G^2 Y_5 \notag
  \\
  &\quad\,
  - 2 \Big\{ \frac{d}{dt} + d H + (9\!-\!d) G \Big\} \{ - \dot{Y}_2 - d H (Y_2 + Y_4) - (9\!-\!d) G (Y_2 + Y_5) \} \Big]. \notag
\end{alignat}
Here we defined $Y_1 = Y_{0i0i}$, $Y_2 = Y_{0m0m}$, $Y_3 = Y_{ijij}$, $Y_4 = Y_{imim}$, $Y_5 = Y_{mnmn}$
for $i,j=1,\cdots,d$ and $m,n=d+1,\cdots,10$.
$Z$ is given by the eq.~(\ref{eq:Zdef}), and the dot represents the time derivative.
The details of the calculations can be found in the appendix~\ref{App:cal}.

\subsection{Classical Solution}

In the following, we will construct the inflationary solution of the eq.~(\ref{eq:MEOMgen}) up to the linear order of $\Gamma$.
Since we solve the eq.~(\ref{eq:MEOMgen}) order by order,
first let us consider the classical equations of motion which are obtained by setting $\Gamma=0$.
Then the equations of motion (\ref{eq:MEOMgen}) are simplified as
\begin{alignat}{3}
  G = \frac{-d(10-d) \pm 3 \sqrt{d(10-d)}}{(10-d)(9-d)} H, \qquad 
  \dot{H}  + \frac{-d \pm 3\sqrt{d(10-d)}}{9-d} H^2 = 0. \label{eq:formergen}
\end{alignat}
From these equations, $H$ and $G$ are solved as
\begin{alignat}{3}
  H(t) &= \frac{H_\text{I}}{\frac{- d \pm 3\sqrt{d(10-d)}}{9-d} H_\text{I} \, t + 1} , \notag
  \\
  G(t) &= \frac{-d(10-d) \pm 3 \sqrt{d(10-d)}}{(10-d)(9-d)} 
  \frac{H_\text{I}}{\frac{-d \pm 3\sqrt{d(10-d)}}{9-d} H_\text{I} \, t + 1}. \label{eq:HandGgen}
\end{alignat}
$H(0) = H_\text{I}$ is an integral constant.
Then $a(t)$ and $b(t)$ are solved as\cite{Kasner:1921zz,Chodos:1979vk,Akune:2006dg}
\begin{alignat}{3}
  a(t) &= a_\text{I} \Big( \frac{-d \pm 3\sqrt{d(10-d)}}{9-d} H_\text{I} \, t + 1 \Big)^{\frac{d \pm 3\sqrt{d(10-d)}}{10d}}, \notag
  \\
  b(t) &= b_\text{I} \Big( \frac{-d \pm 3\sqrt{d(10-d)}}{9-d} H_\text{I} \, t + 1 \Big)^{\frac{10-d \mp 3\sqrt{d(10-d)}}{10(10-d)}}. \label{eq:aandbClgen}
\end{alignat}
where $a_\text{I}$ and $b_\text{I}$ are integral constants.
We assume that the spacetime appears at $t=0$ with $a(0) = a_\text{I}$ and $b(0) = b_\text{I}$.
Since the value in the parentheses should be positive for $0 \leq t$, we choose $0 \leq H_\text{I}$ for the upper sign
and $H_\text{I} \leq 0$ for the lower sign.
Thus, $a(t)$ expands and $b(t)$ shrinks for the upper sign. 
On the other hand, $a(t)$ shrinks and $b(t)$ expands for the lower sign.
Note that, the solution becomes rather special in the case of the upper sign with $d=1$, i.e., 
$a(t) = a_\text{I} (H_\text{I} \, t + 1)$ and $b(t) = b_\text{I}$.

Below we introduce the dimensionless parameter $\tau$ as
\begin{alignat}{3}
  \tau = \frac{- d \pm 3\sqrt{d(10-d)}}{9-d} H_\text{I} \, t + 1. \label{eq:taugen}
\end{alignat}
The range of $\tau$ becomes $1 \leq \tau$.
By using $\tau$, the classical solution is summarized as
\begin{alignat}{3}
  H(\tau) &= \frac{H_\text{I}}{\tau}, \qquad &
  G(\tau) &= \frac{-d(10-d) \pm 3 \sqrt{d(10-d)}}{(10-d)(9-d)} \frac{H_\text{I}}{\tau}, \notag
  \\
  a(\tau) &= a_\text{I} \, \tau^{\frac{d \pm 3\sqrt{d(10-d)}}{10d}}, \qquad &
  b(\tau) &= b_\text{I} \, \tau^{\frac{10-d \mp 3\sqrt{d(10-d)}}{10(10-d)}}. \label{eq:abgen}
\end{alignat}
As mentioned above $a(\tau)$ or $b(\tau)$ expands as $\tau$ increases, but it is different from the inflationary expansion.
This is consistent with the no-go theorem which states that it is impossible to describe accelerated expansion of Universe
in the framework of supergravity theory\cite{Gibbons:1984kp,Maldacena:2000mw,Gibbons:2003gb}.

\subsection{Quantum Solution}

Now we solve the equations of motion (\ref{eq:MEOMgen}) up to the linear order of $\Gamma$.
In order to execute this task, we expand $H$ and $G$ around the classical solution (\ref{eq:abgen}) as
\begin{alignat}{3}
  H(\tau) &= \frac{H_\text{I}}{\tau} + \Gamma H_\text{I}^7 h(\tau), \qquad
  G(\tau) = \frac{-d(10-d) \pm 3 \sqrt{d(10-d)}}{(10-d)(9-d)} \frac{H_\text{I}}{\tau} + \Gamma H_\text{I}^7 g(\tau). \label{eq:HGQgen}
\end{alignat}
$h(\tau)$ and $g(\tau)$ are dimensionless functions with respect to $\tau$.
Then by substituting the eq.~(\ref{eq:HGQgen}) into the eq.~(\ref{eq:MEOMgen}) and 
comparing the linear terms with respect to $\Gamma$,
we obtain equations for $h(\tau)$ and $g(\tau)$. (We employed Mathematica since calculations are straightforward but tedious.)
$g(\tau)$ is expressed in terms of $h(\tau)$ as
\begin{alignat}{3}
  g(\tau) &= \frac{- d(10-d) \pm 3 \sqrt{d(10-d)}}{(10-d)(9-d)} h(\tau) 
  + \frac{5184}{5 (10-d)^4 (9-d)^7 \tau^7} \big\{ \notag
  \\
  &\quad\,
  \!-\! 12 d(10\!-\!d) \big( 659 d^8 \!-\! 64449 d^7 \!+\! 2054460 d^6 \!-\! 31189820 d^5 \!+\! 251678205 d^4 \notag
  \\
  &\quad\,
  \!-\! 1092597255 d^3 \!+\! 2448054900 d^2 \!-\! 2647363500 d \!+\! 590490000 \big) \label{eq:geq}
  \\
  &\quad\,
  \!\pm\! \sqrt{d(10\!-\!d)} \big( 30181 d^9 \!-\! 1625421 d^8 \!+\! 34944195 d^7 \!-\! 375884005 d^6 \!+\! 2011588245 d^5 \notag
  \\
  &\quad\,
  \!-\! 3752608095 d^4 \!-\! 8785306575 d^3 \!+\! 46520278425 d^2 \!-\! 71493576750 d \!+\! 18600435000 \big) \big\}, \notag
\end{alignat}
and the differential equation for $h(\tau)$ is given by
\begin{alignat}{3}
  &0 = \frac{dh(\tau)}{d\tau} + \frac{2}{\tau} h(\tau) + \frac{5 c_h}{\tau^8}, \label{eq:heq}
  \\[0.1cm]
  &c_h \equiv \frac{-2592}{25 (10 \!-\! d)^3 (9 \!-\! d)^6} \big\{ 
  92731 d^9 \!-\! 4614210 d^8 \!+\! 89812929 d^7 \!-\! 824993140 d^6 \!+\! 2943251745 d^5 \notag
  \\
  &\qquad\,
  \!+\! 6230342070 d^4 \!-\! 81640349145 d^3 \!+\! 234351046800 d^2 \!-\! 294211642500 d \!+\! 72630270000 \notag
  \\
  &\qquad\,
  \!\pm\! \sqrt{d(10\!-\!d)} \big( \!-\! 70213 d^8 \!+\! 4636560 d^7 \!-\! 120791847 d^6 \!+\! 1618078810 d^5 \!-\! 12012374835 d^4 \notag
  \\
  &\qquad\,
  \!+\! 49358796660 d^3 \!-\! 106905156345 d^2 \!+\! 111881124450 d \!-\! 23678649000 \big) \big\}. \notag
\end{alignat}
By integrating the above equation, $h(\tau)$ is easily solved. 
And by inserting $h(\tau)$ into the eq.~(\ref{eq:geq}), $g(\tau)$ can be obtained.
The result becomes
\begin{alignat}{3}
  &h(\tau) = \frac{c}{\tau^2} + \frac{c_h}{\tau^7}, \qquad
  g(\tau) = \frac{- d(10-d) \pm 3 \sqrt{d(10-d)}}{(10-d)(9-d)} \frac{c}{\tau^2} + \frac{c_g}{\tau^7}, \label{eq:hgsol}
  \\[0.1cm]
  &c_g \equiv \frac{-2592}{25 (10\!-\!d)^4 (9\!-\!d)^6} 
  \big\{ d(10\!-\!d) \big( 92731 d^8 \!-\! 3648072 d^7 \!+\! 50804481 d^6 \!-\! 251912470 d^5 \notag
  \\
  &\qquad\,
  \!-\! 435418515 d^4 \!+\! 8147315340 d^3 \!-\! 25279230465 d^2 \!+\! 33786853650 d \!-\! 8089713000 \big) \notag
  \\
  &\qquad\,
  \!\pm\! \sqrt{d(10\!-\!d)} \big( 70213 d^9 \!-\! 4683156 d^8 \!+\! 122597463 d^7 \!-\! 1642616950 d^6 \!+\! 12125749755 d^5 \notag
  \\
  &\qquad\,
  \!-\! 49064670000 d^4 \!+\! 102693985785 d^3 \!-\! 99618834150 d^2 \!+\! 8070030000 d \!+\! 3542940000 \big) \big\}. \notag
\end{alignat}
Note that the constant $c$ appears in $h(\tau)$ and $g(\tau)$, but this corresponds to the moduli parameter of the classical solution (\ref{eq:abgen})
up to the linear approximation, and it can be absorbed by shifting $H_\text{I}$.
Therefore it is possible to set $c=0$ without loss of generality.
From the dimensional analysis, it is expected that there are terms of $\Gamma^k H_\text{I}^{6k+1}/ \tau^{6k+1}$ from higher derivative corrections in general.
Since we have poor knowledge about terms for $2\leq k$, we simply consider the region $\Gamma H_\text{I}^6 \ll 1$ and neglect those terms.

Finally $a(\tau)$ and $b(\tau)$ are given like
\begin{alignat}{3}
  a(\tau) &= a_\text{I} \, \tau^{\frac{d \pm 3\sqrt{d(10-d)}}{10d}} 
  \exp \Big[ \frac{d \pm 3\sqrt{d(10-d)}}{60d} c_h \Gamma H_\text{I}^6 \Big( 1 - \frac{1}{\tau^6} \Big) \Big], \notag
  \\
  b(\tau) &= b_\text{I} \, \tau^{\frac{10-d \mp 3\sqrt{d(10-d)}}{10(10-d)}} 
  \exp \Big[ \frac{d \pm 3\sqrt{d(10-d)}}{60d} c_g \Gamma H_\text{I}^6 \Big( 1 - \frac{1}{\tau^6} \Big) \Big], \label{eq:absol}
\end{alignat}
where $a_\text{I}$ and $b_\text{I}$ are initial values of $a(\tau)$ and $b(\tau)$ at $\tau=1$.
The exponential parts come from quantum corrections, and they cause inflation or deflation during $1 \leq \tau < 2$.
Of course, the above expressions are reliable up to $\mathcal{O}((\Gamma H_\text{I}^6)^2)$.
If we take into account $\Gamma^k H_\text{I}^{6k+1}/\tau^{6k+1}$ terms in the eq.~(\ref{eq:HGQgen}), we should add terms like $\Gamma^k H_\text{I}^{6k} (1-1/k^{6k})$ in the argument of the exponentials in the above.
 These terms would also contribute to the inflation or deflation, but as long as we consider the region $\Gamma H_\text{I}^6  \ll 1$,
the leading behaviors of the inflation and deflation are well controlled by the solution (\ref{eq:absol}).

\section{Inflationary Solution in 4 Dimensional Spacetime} \label{sec:Inf4D}

In this section, we consider $4$ dimensional spacetime with $7$ internal directions. 
Namely, we examine the solutions obtained in the previous section by setting $d=3$.
Since we are interested in the inflationary solution in 4 dimensional spacetime, 
we also choose the upper sign in the equations in the previous section. 

\begin{figure}[htb]
\begin{center}
\begin{picture}(240,170)
\put(249,7){$\tau$}
\put(200,140){$\frac{a}{a_\text{I}}$}
\put(200,25){$\frac{b}{b_\text{I}}$}
\includegraphics[width=8.5cm]{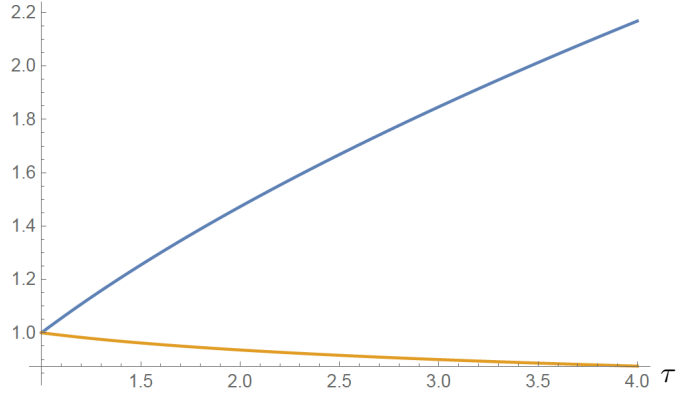}
\end{picture}
\caption{Plots of $a(\tau)/a_\text{I}$ and $b(\tau)/b_\text{I}$ in the eq.~(\ref{eq:ab2}). 
$a(\tau)/a_\text{I}$ is not inflating exponentially.}
\label{fig:classical}
\end{center}
\end{figure}

First, the classical solution (\ref{eq:abgen}) becomes like
\begin{alignat}{3}
  H(\tau) &= \frac{H_\text{I}}{\tau}, \qquad &
  G(\tau) &= \frac{-7 + \sqrt{21}}{14}\frac{H_\text{I}}{\tau}, \notag
  \\
  a(\tau) &= a_\text{I} \, \tau^\frac{1 + \sqrt{21}}{10}, \qquad &
  b(\tau) &= b_\text{I} \, \tau^{-\frac{3\sqrt{21}-7}{70}}. \label{eq:ab2}
\end{alignat}
However, this does not represent inflationary expansion in 3 spatial directions.
The scale factor behaves like $a(t) \sim t^\frac{1 + \sqrt{21}}{10}$ for large $t$,
so it represents rather a radiation dominant era.
The plots of $a(\tau)/a_\text{I}$ and $b(\tau)/b_\text{I}$ are given in fig.~\ref{fig:classical}.

Now we consider the quantum solutions (\ref{eq:hgsol}) and (\ref{eq:absol}).
By setting $d=3$ and choosing the upper sign, $H$ and $G$ are written as
\begin{alignat}{3}
  H(\tau) &= \frac{H_\text{I}}{\tau} + \frac{c_h \Gamma H_\text{I}^7}{\tau^7}, \qquad
  &&c_h = \frac{13824 (477087 \!-\! 97732\sqrt{21})}{8575} \sim 47111, \notag
  \\
  G(\tau) &= \frac{-7 + \sqrt{21}}{14} \frac{H_\text{I}}{\tau} + \frac{c_g \Gamma H_\text{I}^7}{\tau^7}, \qquad
  &&c_g = - \frac{41472 (532196 \!-\! 110451 \sqrt{21})}{60025} \sim -17996. \label{eq:HGsol}
\end{alignat}
And the scale factors $a(\tau)$ and $b(\tau)$ are obtained as
\begin{alignat}{3}
  a(\tau) &= a_\text{I} \, \tau^{\frac{1+\sqrt{21}}{10}} \exp \Big[ \frac{1+\sqrt{21}}{60} c_h \Gamma H_\text{I}^6 
  \Big( 1 - \frac{1}{\tau^6} \Big) \Big], \notag
  \\
  b(\tau) &= b_\text{I} \, \tau^{-\frac{3\sqrt{21}-7}{70}} \exp \Big[ \frac{1+\sqrt{21}}{60} c_g \Gamma H_\text{I}^6 
  \Big( 1 - \frac{1}{\tau^6} \Big) \Big], \label{eq:abQc}
\end{alignat}
where $a_\text{I}$ and $b_\text{I}$ are initial values of $a(\tau)$ and $b(\tau)$ at $\tau=1$.
The exponential parts come from quantum corrections, and
they cause inflation for 3 spatial directions and  deflation for 7 internal directions during $1 \leq \tau < 2$. 
The above expressions are reliable up to $\mathcal{O}((\Gamma H_\text{I}^6)^2)$, but as long as we consider the region $\Gamma H_\text{I}^6 \ll 1$,
the leading behaviors of the inflation or deflation are well controlled by the solution (\ref{eq:abQc}).

\begin{figure}[htb]
\begin{center}
\begin{picture}(240,170)
\put(245,5){$\tau$}
\put(40,145){$\log \frac{a}{a_\text{I}}$}
\put(40,25){$\log \frac{b}{b_\text{I}}$}
\includegraphics[width=8.5cm]{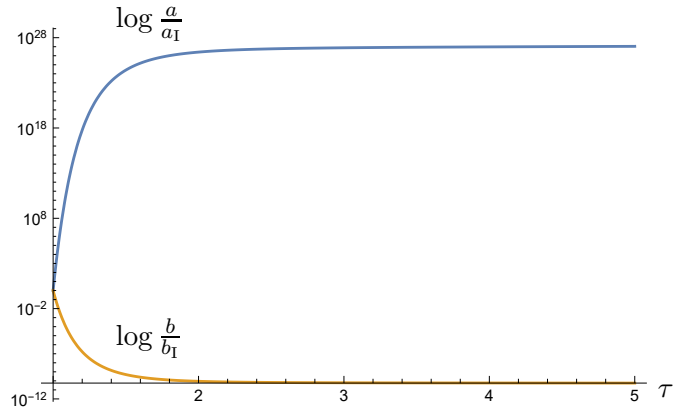}
\end{picture}
\caption{Plots of $\log \big( a(\tau)/a_\text{I} \big)$ and $\log \big( b(\tau)/b_\text{I} \big)$ 
in the eq.~(\ref{eq:abQc}) with $\Gamma H_\text{I}^6 \sim 0.014$.  
The ratio of the scale factor $a(\tau)/a_\text{I}$ in 3 spatial directions increases rapidly during $1 \leq \tau < 2$.}
\label{fig:quantum}
\end{center}
\end{figure}

From the eq.~(\ref{eq:abQc}), the e-folding number $N_\text{e}$ is defined by
\begin{alignat}{3}
  N_\text{e} = \frac{1+\sqrt{21}}{60} c_h \Gamma H_\text{I}^6 \sim 4383 \, \Gamma H_\text{I}^6,
\end{alignat}
and if we set $N_\text{e} = 60$, we obtain $\Gamma H_\text{I}^6 \sim 0.014$.
Then $\frac{1+\sqrt{21}}{60} c_g \Gamma H_\text{I}^6 \sim - 23$ for the deceleration.
The plots of $\log \big( a(\tau)/a_\text{I} \big)$ and $\log \big( b(\tau)/b_\text{I} \big)$ in the eq.~(\ref{eq:abQc}) with $\Gamma H_\text{I}^6 \sim 0.014$ 
are shown in fig.~\ref{fig:quantum}.
The ratio of the scale factor $a(t)/a_\text{I}$ in 3 spatial directions increases rapidly during $1 \leq \tau < 2$.
Especially when $\tau$ is close to $1$, the scale factor expands exponentially with respect to $\tau$.
After the end of the inflation, the expansion rate in 3 spatial directions becomes smaller and behaves like the classical solution 
$a(\tau)/a_\text{I} \sim \tau^{\frac{1+\sqrt{21}}{10}}$.
In other words, the inflation ends naturally when the curvature radius becomes large and quantum corrections are negligible.
On the other hand, the ratio of the scale factor $b(\tau)/b_\text{I}$ in 7 internal directions decreases rapidly during $1 \leq \tau < 2$.
The deflation ends naturally when quantum corrections are negligible and 
the ratio of the scale factor behaves like $b(\tau)/b_\text{I} \sim \tau^{-\frac{3\sqrt{21}-7}{70}}$.

\begin{figure}[htb]
\begin{center}
\begin{picture}(240,170)
\put(245,5){$\tau$}
\put(3,155){$\epsilon$}
\includegraphics[width=8.5cm]{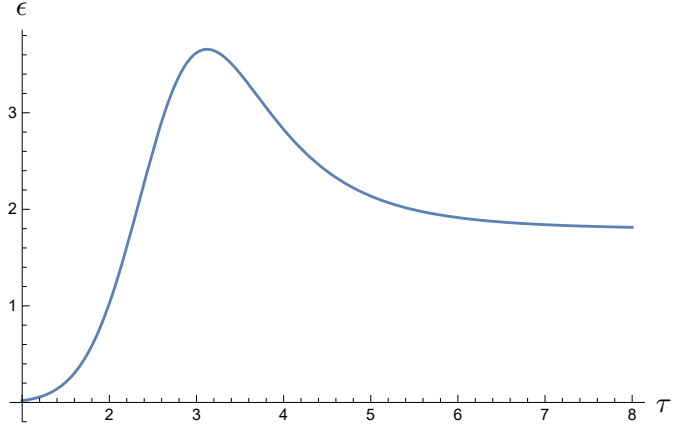}
\end{picture}
\caption{Slow roll parameter $\epsilon$ for $\Gamma H_\text{I}^6 = 0.014$. Inflation ends around $\tau = 2$ since $\epsilon$ becomes almost $1$ there.}
\label{fig:epsilon}
\end{center}
\end{figure}

Finally let us examine the slow roll parameter $\epsilon=-\dot{H}/H^2$, which is one of the important parameters 
for the inflationary cosmology. The inflation lasts during $\epsilon < 1$.
The explicit form of the function is given by
\begin{alignat}{3}
  \epsilon &= \frac{-1+\sqrt{21}}{2} \Big( 1 + \frac{7 c_h \Gamma H_\text{I}^6}{\tau^6} \Big)
  \Big( 1 + \frac{c_h \Gamma H_\text{I}^6}{\tau^6} \Big)^{-2}, \label{eq:slowroll}
\end{alignat}
if we neglect higher order terms of $\mathcal{O}((\Gamma H_\text{I}^6)^2)$ in $H(\tau)$.
The plot of the slow roll parameter (\ref{eq:slowroll}) is shown in fig.~\ref{fig:epsilon}.
From this we see that the inflation ends around $\tau = 2$ because the slow roll parameter $\epsilon$ 
becomes almost $1$ there.

\section{Analyses of $(d+1)$ Dimensional Spacetime} \label{sec:InfanyD}

In the previous section, we obtained inflationary solution in 4 dimensional spacetime.
Then it is natural to ask whether it is true or not in other dimensions.

\begin{figure}[tb]
\begin{center}
\begin{picture}(240,170)
\put(245,3){$\tau$}
\put(-13,168){$\frac{a}{a_\text{I}}$ or $\frac{b}{b_\text{I}}$}
\put(245,155){$d=1$}
\put(245,85){$d=2$}
\put(245,58){$d=3$}
\put(255,35){$\vdots$}
\put(245,17){$d=9$}
\includegraphics[width=8.5cm]{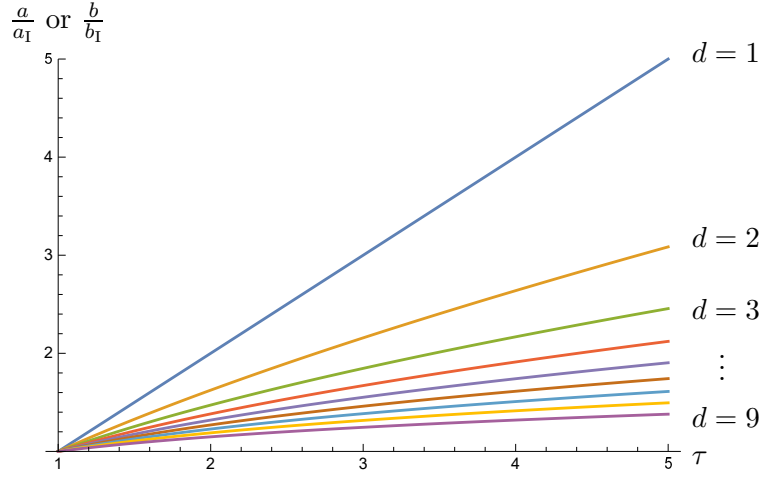}
\end{picture}
\caption{Plots of $a(\tau)/a_\text{I}$ or $b(\tau)/b_\text{I}$ in the eq.~(\ref{eq:clany}). }
\label{fig:clany}
\end{center}
\end{figure}

Below we focus on the case where $d$ dimensional spatial directions are expanding and 
$(10-d)$ internal directions are shrinking.
The classical solutions are given by the eq.~(\ref{eq:abgen}).
For $d=1,2,3,4,5$, the scale factor for the expansion is given by $a(\tau)/a_\text{I}$ with the upper sign.
And for $d=6,7,8,9$, the scale factor for the expansion is expressed by $b(\tau)/b_\text{I}$ with the lower sign.
The explicit forms of the expanding scale factors are written as
\begin{alignat}{3}
  \frac{a(\tau)}{a_\text{I}} &= 
  \begin{cases} 
    \tau \qquad\quad\; (d=1) \\ 
    \tau^{\frac{7}{10}} \qquad\;\, (d=2) \\ 
    \tau^{\frac{1 + \sqrt{21}}{10}} \quad (d=3) \\ 
    \tau^{\frac{2 + 3 \sqrt{6}}{20}} \quad (d=4) \\ 
    \tau^\frac{2}{5} \qquad\;\;\, (d=5) 
  \end{cases}, \qquad
  &\frac{b(\tau)}{b_\text{I}} &= 
  \begin{cases} 
    \tau^{\frac{1 + \sqrt{6}}{10}} \quad\;\;\; (d=6) \\ 
    \tau^{\frac{7 + 3\sqrt{21}}{70}} \quad\, (d=7) \\ 
    \tau^{\frac{1}{4}} \qquad\quad\, (d=8) \\ 
    \tau^{\frac{1}{5}} \qquad\quad\, (d=9) 
  \end{cases}. \label{eq:clany}
\end{alignat}
The plots of the above are shown in fig.~\ref{fig:clany}.
The scale factor is expanding for each case, but it does not represent the inflationary expansion.

\begin{figure}[tb]
\begin{center}
\begin{picture}(240,170)
\put(245,5){$\tau$}
\put(-20,165){$\log \frac{a}{a_\text{I}}$ or $\log \frac{b}{b_\text{I}}$}
\put(200,15){$d=1$}
\put(245,155){$d=2$}
\put(256,144){$\vdots$}
\put(245,135){$d=9$}
\includegraphics[width=8.5cm]{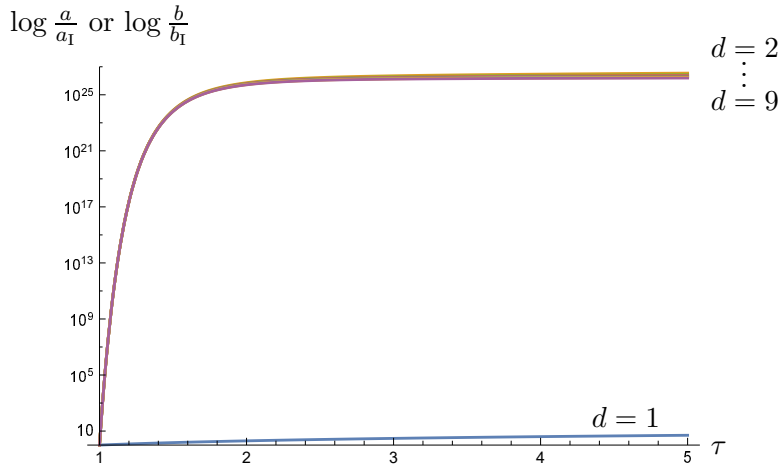}
\end{picture}
\caption{Plots of $\log \big( a(\tau)/a_\text{I} \big)$ and $\log \big( b(\tau)/b_\text{I} \big)$ in the eq.~(\ref{eq:qmany}) 
with choosing $\Gamma H_\text{I}^6$ as the eq.~(\ref{eq:gammaH6}). The inflation occurs except for $d=1$.}
\label{fig:quantumany}
\end{center}
\end{figure}

Now we consider the quantum solutions (\ref{eq:hgsol}) and (\ref{eq:absol}).
For $d=1,2,3,4,5$, the scale factor for the expansion is given by $a(\tau)/a_\text{I}$ with the upper sign.
And for $d=6,7,8,9$, the scale factor for the expansion is expressed by $b(\tau)/b_\text{I}$ with the lower sign.
The explicit expressions for the scale factors are given by
\begin{alignat}{3}
  \frac{a(\tau)}{a_\text{I}} &= 
  \begin{cases} 
    \tau \hspace{8.35cm} (d=1) \\[0.1cm]
    \tau^{\frac{7}{10}} \exp \Big\{ \frac{88134777}{420175} \Gamma H_\text{I}^6 \big( 1 - \frac{1}{\tau^6} \big) \Big\} 
          \hspace{3.3cm} (d=2) \\[0.1cm]
    \tau^{\frac{1 + \sqrt{21}}{10}} \exp \Big\{ \frac{1152 (\!-\! 315057 \!+\! 75871 \sqrt{21})}{8575} 
          \Gamma H_\text{I}^6 \big( 1 - \frac{1}{\tau^6} \big) \Big\} 
          \quad\;\;\, (d=3) \\[0.1cm]
    \tau^{\frac{2 + 3 \sqrt{6}}{20}} \exp \Big\{ \frac{7344 (-430936 + 188879 \sqrt{6})}{15625} 
          \Gamma H_\text{I}^6 \big( 1 - \frac{1}{\tau^6} \big) \Big\} 
          \quad\, (d=4) \\[0.1cm]
    \tau^\frac{2}{5} \exp \Big\{ \frac{862812}{25} \Gamma H_\text{I}^6 \big( 1 - \frac{1}{\tau^6} \big) \Big\} 
          \hspace{3.7cm} (d=5) 
  \end{cases}, \notag
  \\
  \frac{b(\tau)}{b_\text{I}} &= 
  \begin{cases} 
    \tau^{\frac{1 + \sqrt{6}}{10}} \exp \Big\{ \frac{288 (41176038 + 17797607 \sqrt{6})}{78125} 
          \Gamma H_\text{I}^6 \big( 1 - \frac{1}{\tau^6} \big) \Big\} \quad (d=6) \\[0.1cm] 
    \tau^{\frac{7 + 3\sqrt{21}}{70}} \exp \Big\{ \frac{3456 (357455 + 84349 \sqrt{21})}{60025} 
          \Gamma H_\text{I}^6 \big( 1 - \frac{1}{\tau^6} \big) \Big\} \quad\;\, (d=7) \\[0.1cm] 
    \tau^{\frac{1}{4}} \exp \Big\{ \frac{1293111}{200} \Gamma H_\text{I}^6 \big( 1 - \frac{1}{\tau^6} \big) \Big\} 
          \hspace{3.6cm} (d=8) \\[0.1cm] 
    \tau^{\frac{1}{5}} \exp \Big\{ \frac{458631}{400} \Gamma H_\text{I}^6 \big( 1 - \frac{1}{\tau^6} \big) \Big\} 
          \hspace{3.75cm} (d=9) 
  \end{cases}. \label{eq:qmany}
\end{alignat}
Note that the definitions of $\tau$ depend on $d$. Also parameters $H_\text{I}$ can be chosen arbitrary for each $d$.
The e-folding numbers are defined by the coefficients of $( 1 - \frac{1}{\tau^6} )$ in the exponentials,
and we will fix $\Gamma H_\text{I}^6$ for $d=2,\cdots,9$ by setting the e-folding numbers to be 60.
That is, we set 
\begin{alignat}{3}
  \Gamma H_\text{I}^6 \sim 0.29,\; 0.014,\; 0.0040,\; 0.0017,\; 0.00019,\; 0.0014,\; 0.0093,\; 0.052, \label{eq:gammaH6}
\end{alignat}
for $d=2, \cdots, 9$. Since the above values are smaller than 1, we may neglect the higher order corrections.
The plots of $\log \big( a(\tau)/a_\text{I} \big)$ and $\log \big( b(\tau)/b_\text{I} \big)$ in the eq.~(\ref{eq:qmany}) 
by using the above values of $\Gamma H_\text{I}^6$ are shown in fig.~\ref{fig:quantumany}.
The inflation does not occur for $d=1$, and for $d=2,\cdots,9$ the inflation occurs during $1 \leq \tau < 2$.
Thus the inflationary scenario is universal if we introduce the quantum corrections to the classical gravity.

\section{Conclusion and Discussion}\label{sec:con}

In this paper we have investigated inflationary cosmology in M-theory.
In the low energy limit, the M-theory is approximated by the 11 dimensional supergravity.
If we take into account quantum corrections, however, the 11 dimensional supergravity is corrected by quartic terms of the Weyl tensor at the leading order.
We have divided 10 spatial directions into $d$ spatial ones and $(10-d)$ internal ones with different scale factors, 
and investigated the effects of the quantum corrections to the cosmological evolution.

In the low energy limit, the solution represents that $d$ spatial directions slowly expand and $(10-d)$ internal directions gradually shrink.
This qualitative aspects can be understood naively by the dimensional analysis. 
Namely the Hubble parameter behaves like $W \sim H^2 \sim \tau^{-2}$, so the scale factor does like $\log a \sim \log \tau$.
This does not match with the inflationary cosmology, and it is consistent with the no-go theorem
which states that it is difficult to realize accelerated expansion of Universe in the low energy limit of the string theory\cite{Gibbons:1984kp,Maldacena:2000mw,Gibbons:2003gb}.

If we correct the 11 dimensional supergravity by adding $W^4$ terms, the Hubble parameter is modified like
$H^2 \sim  \tau^{-2} (1 + \Gamma \tau^{-6})^2$, so the scale factor behaves like $\log a \sim \log \tau + \Gamma \tau^{-6}$ up to the linear order of $\Gamma$.
Then the quantum corrections give exponential expansion of the scale factor when $\tau$ is close to 1.
And the inflation naturally ends when the quantum corrections are negligible compared to the classical supergravity part.
This is also true for the internal directions.
The scale factor for the internal directions rapidly deflates around $\tau=1$, and changes to shrink gradually 
when the quantum corrections are negligible compared to the classical part.
The explicit solutions of the scale factors are given by the eq.~(\ref{eq:absol}).

Especially for the case of $d=3$, we showed the plots of $a(\tau)$ and $b(\tau)$ in the fig.~\ref{fig:quantum} by fixing $\Gamma H_{\text{I}}^6 \sim 0.014$,
which is determined by setting the e-folding number $N_\text{e} = 60$.
The deceleration of the internal directions are also fixed and the e-folding number is estimated as $-23$.
We plotted the slow roll parameter $\epsilon$ in the fig.~\ref{fig:epsilon}, and confirmed that $\epsilon < 1$ when $1 \leq \tau<2$.
For the case of $d=2,\cdots,9$, the qualitative feature of the scale factors are the same as those for the case of $d=3$.
Thus we conclude that the inflation is universal if we take into account the quantum corrections in the M-theory.
Notice that $d=1$ and $d=10$ are rather special and do not receive quantum corrections, hence there are no inflationary solutions.

As a future work, it is important to examine other cosmological parameters for the case of $d=3$.
We will introduce the fluctuations around the background metric, and evaluate the spectral index $n_s$ and the tensor to scalar ratio $r$
to compare them with observations. 
It is necessary to compare our results around $\tau \sim 1$ with those in the refs.~\cite{Ohta:2004wk,Maeda:2004vm,Maeda:2004hu,Akune:2006dg,Maeda:2011zn}.
If the results are compatible with each other, it is possible to seek the inflationary solutions by using simpler metric ansatz of $e^{Ht}$ and $e^{Gt}$.
It is also interesting to apply the analyses of this paper 
to the heterotic superstring theory with nontrivial internal space, which contains $R^2$ corrections\cite{Brandle:2000qp},
and reveal several problems in string cosmology\cite{Antoniadis:2016avv}.

\section*{Acknowledgement}
The authors would like to thank Hiroyuki Abe, Takanori Fujiwara, Keisuke Izumi, Sergei V. Ketov, Hiroyuki Kitamoto, Kazunori Kohri, Keiichi Maeda, 
Shun'ya Mizoguchi, Nobuyoshi Ohta, Makoto Sakaguchi and Yutaka Sakamura for useful discussions.
We would also like to thank the Yukawa Institute for Theoretical Physics
at Kyoto University for hospitality during the workshop YITP-W-17-08 ``Strings and Fields 2017" 
and YITP-T-18-04 ``New Frontiers in String Theory 2018", where part of this work was carried out.
This work was partially supported by Japan Society for the Promotion of Science, Grant-in-Aid for Scientific Research (C) Grant Number JP17K05405.

\appendix
\section{The derivation of eq.~(\ref{eq:MEOMgen})} \label{App:cal}

In this appendix, we derive the eq.~(\ref{eq:MEOMgen}).
First, it is straightforward to evaluate the Weyl tensor by using the ansatz (\ref{eq:metricgen}).
We define $W_1 = W_{0i0i}$, $W_2 = W_{0m0m}$, $W_3 = W_{ijij}$, $W_4 = Y_{imim}$, $W_5 = Y_{mnmn}$
for $i,j=1,\cdots,d$ and $m,n=d+1,\cdots,10$, and those are written as
\begin{alignat}{3}
  W_1 &= \frac{4(d\!-\!10)}{45} \dot{H} \!+\! \frac{4(10\!-\!d)}{45} \dot{G}
  \!-\! \frac{(10\!-\!d)(9\!-\!d)}{90} H^2 \!+\! \frac{(10\!-\!d)(d\!-\!1)}{90} G^2 \!+\! \frac{(5\!-\!d)(10\!-\!d)}{45} HG, \notag
  \\
  W_2 &= \frac{4d}{45} \dot{H} \!-\! \frac{4d}{45} \dot{G}
  \!+\! \frac{d(9\!-\!d)}{90} H^2 \!-\! \frac{d(d\!-\!1)}{90} G^2 \!-\! \frac{d(5\!-\!d)}{45} HG, \notag
  \\
  W_3 &= \frac{d\!-\!10}{45} \dot{H} \!+\! \frac{10\!-\!d}{45} \dot{G}
  \!+\! \frac{(10\!-\!d)(9\!-\!d)}{90} H^2 \!+\! \frac{(10\!-\!d)(11\!-\!d)}{90} G^2
  \!-\! \frac{(10\!-\!d)^2}{45} HG,
  \\
  W_4 &= \frac{d\!-\!5}{45} \dot{H} \!+\! \frac{5\!-\!d}{45} \dot{G}
  \!-\! \frac{d(9\!-\!d)}{90} H^2 \!-\! \frac{(d\!-\!1)(10\!-\!d)}{90} G^2 \!-\! \frac{d^2-10d+5}{45} HG, \notag
  \\
  W_5 &= \frac{d}{45} \dot{H} \!-\! \frac{d}{45} \dot{G}
  \!+\! \frac{d(d\!+\!1)}{90} H^2 \!+\! \frac{d(d\!-\!1)}{90} G^2 \!-\! \frac{d^2}{45} HG. \notag
\end{alignat}
Of course, $W_1$ and $W_2$ are exchanged each other if we exchange $H$ with $G$ and $d$ with $(10-d)$.
This is also true for $W_3$ and $W_5$.

Second, let us evaluate $Z = t_8 t_8 W^4 - \frac{1}{4!} \epsilon_{11} \epsilon_{11} W^4$.
Each term in the $Z$ is calculated as follows.
Note that $i,j$ are summed from $1$ to $d$, and $m,n$ are summed from $d+1$ to $10$,
and $a,b,c,d,e,f,g,h$ are summed from $0$ to $10$ below.
\begin{alignat}{3}
  &W_{abcd} W^{abcd} W_{efgh} W^{efgh} \notag
  \\
  &= 4 W_{abab} W^{abab} W_{efef} W^{efef} \notag
  \\
  &= 4 \big(2 W_{0i0i}^2 + 2 W_{0m0m}^2 + W_{ijij}^2 + 2 W_{imim}^2 + W_{mnmn}^2 \big)^2 \notag
  \\
  &= 16 d^2 W_1^4 + 16 (10\!-\!d)^2 W_2^4 + 4 d^2(d\!-\!1)^2 W_3^4
  + 16 d^2 (10\!-\!d)^2 W_4^4 + 4 (10\!-\!d)^2 (9\!-\!d)^2 W_5^4 \notag
  \\
  &\quad
  + 32 d (10\!-\!d) W_1^2 W_2^2 + 16 d^2 (d\!-\!1) W_1^2 W_3^2 + 32 d^2 (10\!-\!d) W_1^2 W_4^2 + 16 d (10\!-\!d)(9\!-\!d) W_1^2 W_5^2 \notag
  \\
  &\quad
  + 16 d(d\!-\!1) (10\!-\!d) W_2^2 W_3^2 + 32 d (10\!-\!d)^2 W_2^2 W_4^2 + 16 (10\!-\!d)^2 (9\!-\!d) W_2^2 W_5^2 
  \\
  &\quad
  + 16 d^2(d\!-\!1) (10\!-\!d) W_3^2 W_4^2 + 8 d(d\!-\!1) (10\!-\!d)(9\!-\!d) W_3^2 W_5^2 + 16 d(10\!-\!d)^2(9\!-\!d) W_4^2 W_5^2. \notag
\end{alignat}
\vspace{-1cm}
\begin{alignat}{3}
  &W_{abcd} W^{aefg} W^{bcdh} W_{efgh} \notag
  \\
  &=W_{abab} W^{abab} W_{aeae} W^{aeae} \notag
  \\
  &= W_{0i0i}^2 W_{0e0e}^2 + W_{0i0i}^2 W_{ieie}^2 + W_{0m0m}^2 W_{0e0e}^2 + W_{0m0m}^2 W_{meme}^2 \notag
  \\
  &\quad
  + W_{ijij}^2 W_{ieie}^2 + W_{imim}^2 W_{ieie}^2 + W_{imim}^2 W_{meme}^2 + W_{mnmn}^2 W_{meme}^2 \notag
  \\
  &= d(d\!+\!1) W_1^4 + (10\!-\!d)(11\!-\!d) W_2^4 + d(d\!-\!1)^2 W_3^4 + 10d(10\!-\!d) W_4^4 + (10\!-\!d)(9\!-\!d)^2 W_5^4 \notag
  \\
  &\quad
  + 2 d(10\!-\!d) W_1^2 W_2^2 + 2 d(d\!-\!1) W_1^2 W_3^2 + 2 d(10\!-\!d) W_1^2 W_4^2 + 2 d(10\!-\!d) W_2^2 W_4^2
  \\
  &\quad
  + 2 (10\!-\!d)(9\!-\!d) W_2^2 W_5^2 + 2 d(d\!-\!1)(10\!-\!d) W_3^2 W_4^2 + 2 d(10\!-\!d)(9\!-\!d) W_4^2 W_5^2. \notag
\end{alignat}
\vspace{-1cm}
\begin{alignat}{3}
  &W_{abcd} W^{abef} W^{cdgh} W_{efgh} \notag
  \\
  &= 8 W_{abab} W_{abab} W^{abab} W^{abab} \notag
  \\
  &= 16 W_{0i0i}^4 + 16 W_{0m0m}^4 + 8 W_{ijij}^4  + 16 W_{imim}^4 + 8 W_{mnmn}^4 \notag
  \\
  &= 16 d W_1^4 + 16 (10\!-\!d) W_2^4 + 8 d(d\!-\!1) W_3^4  + 16 d(10\!-\!d) W_4^4 + 8 (10\!-\!d)(9\!-\!d) W_5^4.
\end{alignat}
\vspace{-1cm}
\begin{alignat}{3}
  &W_{acbd} W^{aebf} W^{cgdh} W_{egfh} \notag
  \\
  &= W_{acac} W^{aeae} W^{cgcg} W_{egeg} + W_{acac} W_{acac} W^{acac} W^{acac} \notag
  \\
  &= 2 W_{0i0i} W_{0e0e} W_{igig} W_{egeg} + 2 W_{0i0i}^4 
  + 2 W_{0m0m} W_{0e0e} W_{mgmg} W_{egeg} + 2 W_{0m0m}^4 \notag
  \\
  &\quad
  + W_{ijij} W_{ieie} W_{jgjg} W_{egeg} + W_{ijij}^4
  + 2 W_{imim} W_{ieie} W_{mgmg} W_{egeg} + 2 W_{imim}^4 \notag
  \\
  &\quad
  + W_{mnmn} W_{meme} W_{ngng} W_{egeg} + W_{mnmn}^4 \notag
  \\
  &= 2 d(d\!+\!1) W_1^4 + 2 (10\!-\!d)(11\!-\!d) W_2^4 + d(d\!-\!1)(d^2\!-\!3d\!+\!4) W_3^4 \notag
  \\
  &\quad
  + 2 d(10\!-\!d)(1\!+\!10d\!-\!d^2) W_4^4 + (10\!-\!d)(9\!-\!d)(d^2\!-\!17d\!+\!74) W_5^4 + 4 d(10\!-\!d) W_1^2 W_2^2 
  \\
  &\quad
  + 4 d (d\!-\!1)^2 W_1^2 W_3^2 + 4 d^2(10\!-\!d) W_1^2 W_4^2 + 4 d(10\!-\!d)^2 W_2^2 W_4^2 + 4 (10\!-\!d)(9\!-\!d)^2 W_2^2 W_5^2 \notag
  \\
  &\quad
  + 4 d(d\!-\!1)^2(10\!-\!d) W_3^2 W_4^2 + 4 d(10\!-\!d)(9\!-\!d)^2 W_4^2 W_5^2 + 8 d(d\!-\!1)(10\!-\!d) W_1 W_2 W_3 W_4 \notag
  \\
  &\quad
  + 8 d(10\!-\!d)(9\!-\!d) W_1 W_2 W_4 W_5 + 4 d(d\!-\!1)(10\!-\!d)(9\!-\!d) W_3 W_4^2 W_5. \notag
\end{alignat}
\vspace{-1cm}
\begin{alignat}{3}
  &W_{abcd} W^{aefg} W^b{}_{ef}{}^h W^{cd}{}_{gh} + W_{abcd} W^{aefg} W^b{}_{fe}{}^h W^{cd}{}_{gh} \notag
  \\
  &= 4 W_{abab} W^{aeae} W^b{}_e{}^b{}_e W^{ab}{}_{ab} - 2 W_{abab} W^{abab} W_a{}^{ba}{}_b W^{ab}{}_{ab} \notag
  \\
  &=\! - 8 W_{0i0i}^2 W_{0e0e} W_{ieie} - 4 W_{0i0i}^4
  - 8 W_{0m0m}^2 W_{0e0e} W_{meme} - 4 W_{0m0m}^4 + 4 W_{ijij}^2 W_{ieie} W_{jeje} \notag
  \\
  &\quad
  - 2 W_{ijij}^4 + 8 W_{imim}^2 W_{ieie} W_{meme} - 4 W_{imim}^4 
  + 4 W_{mnmn}^2 W_{meme} W_{nene} - 2 W_{mnmn}^4 \notag
  \\
  &=\! - 4 d W_1^4 - 4 (10\!-\!d) W_2^4 + 2 d(d\!-\!1)(2d\!-\!5) W_3^4 - 4 d(10\!-\!d) W_4^4 \notag
  \\
  &\quad
  + 2 (10\!-\!d)(9\!-\!d)(15\!-\!2d) W_5^4 - 8 d(d\!-\!1) W_1^3 W_3 - 8 (10\!-\!d)(9\!-\!d) W_2^3 W_5 \notag
  \\
  &\quad
  + 8 d(d\!-\!1)(10\!-\!d) W_3 W_4^3 + 8 d(10\!-\!d)(9\!-\!d) W_4^3 W_5 + 4 d(d\!-\!1) W_1^2 W_3^2 
  \\
  &\quad
  + 4 (10\!-\!d)(9\!-\!d) W_2^2 W_5^2 + 4 d(d\!-\!1)(10\!-\!d) W_3^2 W_4^2 + 4 d(10\!-\!d)(9\!-\!d) W_4^2 W_5^2 \notag
  \\
  &\quad
  - 8 d(10\!-\!d) W_1^2 W_2 W_4 - 8 d(10\!-\!d) W_1 W_2^2 W_4 + 8 d(10\!-\!d) W_1 W_2 W_4^2. \notag
\end{alignat}
Combining the above results, $Z$ which is defined by the eq.~(\ref{eq:Zdef}) is obtained as
\begin{alignat}{3}
  Z &= 24 \big\{ - 16 d(d\!-\!4) W_1^4 - 16 (10\!-\!d)(6\!-\!d) W_2^4 + 4 d(d\!-\!1)(5d^2 \!-\! 45 d \!+\! 76) W_3^4 \notag
  \\
  &\quad
  - 16 d (10\!-\!d) (3 d^2 \!-\! 30d \!+\! 32) W_4^4 + 4 (10\!-\!d) (9\!-\!d) (5 d^2 \!-\! 55d \!+\! 126) W_5^4 \notag
  \\
  &\quad
  - 32 d (10\!-\!d) W_1^2 W_2^2 + 16 d (d\!-\!1) (5d \!-\! 16) W_1^2 W_3^2 + 32 d (10\!-\!d) (3d \!-\! 4) W_1^2 W_4^2 \notag
  \\
  &\quad
  + 16 d (10\!-\!d)(9\!-\!d) W_1^2 W_5^2 + 16 d(d\!-\!1) (10\!-\!d) W_2^2 W_3^2 - 32 d (10\!-\!d) (3d \!-\!26) W_2^2 W_4^2 \notag
  \\
  &\quad
  - 16 (10\!-\!d) (9\!-\!d) (5d \!-\! 34) W_2^2 W_5^2 + 16 d(d\!-\!1)(10\!-\!d) (5d \!-\! 16) W_3^2 W_4^2 \notag
  \\
  &\quad
  + 8 d(d\!-\!1) (10\!-\!d)(9\!-\!d) W_3^2 W_5^2 - 16 d(10\!-\!d)(9\!-\!d) (5d \!-\! 34) W_4^2 W_5^2 \notag
  \\
  &\quad
  + 128 d(d\!-\!1)(10\!-\!d) W_1 W_2 W_3 W_4 + 128 d(10\!-\!d)(9\!-\!d) W_1 W_2 W_4 W_5 \label{eq:Zcal}
  \\
  &\quad
  + 64 d(d\!-\!1)(10\!-\!d)(9\!-\!d) W_3 W_4^2 W_5 + 128 d(d\!-\!1) W_1^3 W_3 \notag
  \\
  &\quad
  + 128 (10\!-\!d)(9\!-\!d) W_2^3 W_5 - 128 d(d\!-\!1)(10\!-\!d) W_3 W_4^3 - 128 d(10\!-\!d)(9\!-\!d) W_4^3 W_5 \notag
  \\
  &\quad
  + 128 d(10\!-\!d) W_1^2 W_2 W_4 + 128 d(10\!-\!d) W_1 W_2^2 W_4 - 128 d(10\!-\!d) W_1 W_2 W_4^2 \big\}. \notag
\end{alignat}

Third, let us evaluate $X_{abcd}$ which is defined by the eq.~(\ref{eq:Xdef}). 
Since the Weyl tensor $W_{abcd}$ is nonzero only when $(a,b) = (c,d)$ or $(a,b) = (d,c)$, $X_{abcd}$ is simplified as follows. 
\begin{alignat}{3}
  X_{abcd} &= 192 \big\{ W_{abcd} W_{efef} W^{efef} - 8 W_{abcd} (W_{afaf} W^{afaf} \!+\! W_{bfbf} W^{bfbf}) \notag
  \\
  &\quad\,
  + 4 W_{abcd} (3 W_{abab} W^{abab} \!+\! W_{acac} W^{acac} \!+\! W_{bcbc} W^{bcbc}) \notag
  \\
  &\quad\,
  + 4 (W_{aece} W_{bfdf} \!-\! W_{bece} W_{afdf}) W^{efef} - 8 W_{abcd} W_{af}{}^{af} W_{bf}{}^{bf} \label{eq:Xdef2}
  \\
  &\quad\,
  - 4 (W^{ag}{}_{ag} \!+\! W^{bg}{}_{bg}) (W_{agcg} W_{gbgd} \!-\! W_{agdg} W_{gbgc}) \big\}. \notag
\end{alignat}
In the above, we may also use the relations $W_{abcd} W_{acac}=W_{abcd} W_{bdbd}$ or $W_{abcd} W_{abab}=W_{abcd} W_{cdcd}$. 
From the eq.~(\ref{eq:Xdef2}), we see that nonzero components of $X_{abcd}$ are 
$X_{0i0i}$, $X_{0m0m}$, $X_{ijij}$, $X_{imim}$ and $X_{mnmn}$.
Then these are calculated as follows.
\begin{alignat}{3}
  X_{0i0i} &= 192 \big\{ W_{0i0i} W_{efef}^2 - 8 W_{0i0i} (W_{0f0f}^2 \!+\! W_{ifif}^2) + 16 W_{0i0i}^3 + 4 W_{0e0e} W_{ifif} W_{efef} \notag
  \\
  &\quad
  + 8 W_{0i0i} W_{0f0f} W_{ifif} - 4 (- W_{0g0g} \!+\! W_{igig}) W_{0g0g} W_{gigi} \big\} \notag
  \\
  &= 192 \big[ 2 (4 \!-\! d) W_1^3 - 2 (10\!-\!d) W_1 W_2^2 + 12 (d\!-\!1) W_1^2 W_3 + (5d \!-\! 16)(d\!-\!1) W_1 W_3^2 \notag
  \\
  &\quad
  + 2 (3d\!-\!4)(10\!-\!d) W_1 W_4^2 + (10\!-\!d)(9\!-\!d) W_1 W_5^2 + 4 (10\!-\!d) W_2^2 W_4 - 4 (10\!-\!d) W_2 W_4^2 \notag
  \\
  &\quad
  + 8 (10\!-\!d) W_1 W_2 W_4 + 4 (d\!-\!1)(10\!-\!d) W_2 W_3 W_4 + 4 (10\!-\!d)(9\!-\!d) W_2 W_4 W_5 \big]. 
\end{alignat}
\vspace{-1cm}
\begin{alignat}{3}
  X_{0m0m} &= 192 \big\{ W_{0m0m} W_{efef}^2 - 8 W_{0m0m} (W_{0f0f}^2 \!+\! W_{mfmf}^2) + 16 W_{0m0m}^3 + 4 W_{0e0e} W_{mfmf} W_{efef} \notag
  \\
  &\quad
  + 8 W_{0m0m} W_{0f0f} W_{mfmf} - 4 (- W_{0g0g} \!+\! W_{mgmg}) W_{0g0g} W_{gmgm} \big\} \notag
  \\
  &= 192 \big[ 2 (d\!-\!6) W_2^3 - 2 d W_1^2 W_2 + 4 d W_1^2 W_4 - 4 d W_1 W_4^2 + d(d\!-\!1) W_2 W_3^2 \notag
  \\
  &\quad
  + 2 d(26\!-\!3d) W_2 W_4^2 + 12 (9\!-\!d) W_2^2 W_5 + (34-5d)(9\!-\!d) W_2 W_5^2 \notag
  \\
  &\quad
  + 8 d W_1 W_2 W_4 + 4 d(d\!-\!1) W_1 W_3 W_4 + 4 d(9\!-\!d) W_1 W_4 W_5 \big]. 
\end{alignat}
\vspace{-1cm}
\begin{alignat}{3}
  X_{ijij} &= 192 \big\{ W_{ijij} W_{efef}^2 - 8 W_{ijij} (W_{ifif}^2 \!+\! W_{jfjf}^2) + 16 W_{ijij}^3 + 4 W_{ieie} W_{jfjf} W_{efef} \notag
  \\
  &\quad\,
  - 8 W_{ijij} W_{ifif} W_{jfjf} - 4 (W^{ig}{}_{ig} \!+\! W^{jg}{}_{jg}) W_{igig} W_{gjgj} \big\} \notag
  \\
  &= 192 \big[ 8 W_1^3 + (5d^2 \!-\! 45d \!+\! 76) W_3^3 - 8 (10\!-\!d) W_4^3 + 2 (5d \!-\! 16) W_1^2 W_3 \notag
  \\
  &\quad\,
  + 2 (10\!-\!d) W_2^2 W_3 + 2 (5d\!-\!16)(10\!-\!d) W_3 W_4^2 + (10\!-\!d)(9\!-\!d) W_3 W_5^2 
  \\
  &\quad\,
  + 4 (10\!-\!d)(9\!-\!d) W_4^2 W_5 + 8 (10\!-\!d) W_1 W_2 W_4 \big]. \notag
\end{alignat}
\vspace{-1cm}
\begin{alignat}{3}
  X_{imim} &= 192 \big\{ W_{imim} W_{efef}^2 - 8 W_{imim} (W_{ifif}^2 \!+\! W_{mfmf}^2) + 16 W_{imim}^3 + 4 W_{ieie} W_{mfmf} W_{efef} \notag
  \\
  &\quad\,
  - 8 W_{imim} W_{ifif} W_{mfmf} - 4 (W^{ig}{}_{ig} \!+\! W^{mg}{}_{mg}) W_{igig} W_{gmgm} \big\} \notag
  \\
  &= 192 \big[ 2 (\!- 3d^2 \!+\! 30d \!-\! 32) W_4^3 + 4 W_1^2 W_2 + 4 W_1 W_2^2 + 2 (3d \!-\! 4) W_1^2 W_4 \notag
  \\
  &\quad\,
  + 2 (26\!-\!3d) W_2^2 W_4 + (5d\!-\!16)(d\!-\!1) W_3^2 W_4 - 12 (d\!-\!1) W_3 W_4^2
  \\
  &\quad\,
  - 12 (9\!-\!d) W_4^2 W_5 - (5d\!-\!34)(9\!-\!d) W_4 W_5^2 + 4 (d\!-\!1) W_1 W_2 W_3 - 8 W_1 W_2 W_4 \notag
  \\
  &\quad\,
  + 4 (9\!-\!d) W_1 W_2 W_5 + 4 (d\!-\!1)(9\!-\!d) W_3 W_4 W_5 \big]. \notag
\end{alignat}
\vspace{-1cm}
\begin{alignat}{3}
  X_{mnmn} &= 192 \big\{ W_{mnmn} W_{efef}^2 - 8 W_{mnmn} (W_{mfmf}^2 \!+\! W_{nfnf}^2) + 16 W_{mnmn}^3 + 4 W_{meme} W_{nfnf} W_{efef} \notag
  \\
  &\quad\,
  - 8 W_{mnmn} W_{mfmf} W_{nfnf} - 4 (W^{mg}{}_{mg} \!+\! W^{ng}{}_{ng}) W_{mgmg} W_{gngn} \big\} \notag
  \\
  &= 192 \big[ 8 W_2^3 - 8 d W_4^3 + (5d^2 \!-\! 55d \!+\! 126) W_5^3 + 2 d W_1^2 W_5 + 2 (34\!-\!5d) W_2^2 W_5 \notag
  \\
  &\quad\,
  + 4 d(d\!-\!1) W_3 W_4^2 + d(d\!-\!1) W_3^2 W_5 + 2 d(34\!-\!5d) W_4^2 W_5 + 8 d W_1 W_2 W_4 \big]. 
\end{alignat}
$X_{ac} = X_a{}^d{}_{cd}$ is simplified as 
\begin{alignat}{3}
  X_{ac} &= 2304 \big( - W_a{}^d{}_{cd} W_{dfdf} W^{dfdf} + 2 W_{adcd} W_c{}^{d}{}_{cd} W^{cdcd} 
  - W_{adcd} W^a{}_{faf} W^{dfdf} \big), 
\end{alignat}
and nonzero components are calculated as follows.
\begin{alignat}{3}
  X_{00} &= 2304 \big( - W_{0d0d} W_{dfdf}^2 + 2 W_{0d0d}^3 + W_{0d0d} W_{0f0f} W^{dfdf} \big) \notag
  \\
  &= 2304 \big[ d W_1^3 + (10\!-\!d) W_2^3 + d(d\!-\!1) W_1^2 W_3 - d(d\!-\!1) W_1 W_3^2 - d(10\!-\!d) W_1 W_4^2 
  \\
  &\quad
  - d(10\!-\!d) W_2 W_4^2 - (10\!-\!d)(9\!-\!d) W_2 W_5^2 + (10\!-\!d)(9\!-\!d) W_2^2 W_5 
  + 2 d(10\!-\!d) W_1 W_2 W_4 \big], \notag
  \\
  X_{ii} &= 2304 \big( - W_i{}^d{}_{id} W_{dfdf}^2 + 2 W_i{}^{d}{}_{id} W_{idid}^2 - W_{idid} W_{ifif} W_{dfdf} \big) \notag
  \\
  &= 2304 \big[ (d\!-\!2) W_1^3 - (d\!-\!1)(2d\!-\!5) W_3^3 - (d\!-\!2)(10\!-\!d) W_4^3 \notag
  \\
  &\quad
  + (10\!-\!d) W_1 W_2^2 - 3 (d\!-\!1) W_1^2 W_3 - (10\!-\!d) W_2^2 W_4 - 3 (d\!-\!1)(10\!-\!d) W_3 W_4^2 
  \\
  &\quad
  - (10\!-\!d)(9\!-\!d) W_4^2 W_5 - (10\!-\!d)(9\!-\!d) W_4 W_5^2 - 2 (10\!-\!d) W_1 W_2 W_4 \big], \notag
  \\
  X_{mm} &= 2304 \big( - W_m{}^d{}_{md} W_{dfdf}^2 + 2  W_m{}^{d}{}_{md} W_{mdmd}^2- W_{mdmd} W_{mfmf} W_{dfdf} \big) \notag
  \\
  &= 2304 \big[ (8\!-\!d) W_2^3 - d(8\!-\!d) W_4^3 - (15\!-\!2d)(9\!-\!d) W_5^3 \notag
  \\
  &\quad
  + d W_1^2 W_2 - d W_1^2 W_4 - 3 (9\!-\!d) W_2^2 W_5 - d(d\!-\!1) W_3^2 W_4 - d(d\!-\!1) W_3 W_4^2 
  \\
  &\quad
  - 3 d(9\!-\!d) W_4^2 W_5 - 2 d W_1 W_2 W_4 \big]. \notag
\end{alignat}
$X = X^a{}_a$ is evaluated as
\begin{alignat}{3}
  X &= 96 \big( 12 W^{ad}{}_{ef} W_{adgh} W^{efgh} - 24 W_{aedg} W^a{}_{f}{}^d{}_{h} W^{efgh} \big) \notag
  \\
  &= 1152 \big( 4 W^{ef}{}_{ef} W_{efef}^2 - 2 W_{aeae} W_{afaf} W_{efef} \big) \notag
  \\
  &= 2304 \big[ - 4 d W_1^3 - 4 (10\!-\!d) W_2^3 + 2 d(d\!-\!1) W_3^3 + 4 d(10\!-\!d) W_4^3 + 2 (10\!-\!d)(9\!-\!d) W_5^3 \notag
  \\
  &\quad
  - 2 W_{0i0i} W_{0f0f} W_{ifif} - 2 W_{0m0m} W_{0f0f} W_{mfmf} - W_{ijij} W_{ifif} W_{jfjf} \notag
  \\
  &\quad
  - 2 W_{imim} W_{ifif} W_{mfmf} - W_{mnmn} W_{mfmf} W_{nfnf} \big] \notag
  \\
  &= 2304 \big[ - 4 d W_1^3 - 4 (10\!-\!d) W_2^3 - d(d\!-\!1)(d\!-\!4) W_3^3 + 4 d(10\!-\!d) W_4^3 \notag
  \\
  &\quad
  - (10\!-\!d)(9\!-\!d)(6\!-\!d) W_5^3 - 3 d(d\!-\!1) W_1^2 W_3 - 3 (10\!-\!d)(9\!-\!d) W_2^2 W_5 
  \\
  &\quad
  - 3 d(d\!-\!1)(10\!-\!d) W_3 W_4^2 - 3 d(10\!-\!d)(9\!-\!d) W_4^2 W_5 - 6 d(10\!-\!d) W_1 W_2 W_4 \big]. \notag
\end{alignat}

Forth, let us evaluate $Y_{abcd}$ which is defined by the eq.~(\ref{eq:Ydef}).
There are 5 kinds of nonzero components, so we define them as 
$Y_1 = Y_{0i0i}$, $Y_2 = Y_{0m0m}$, $Y_3 = Y_{ijij}$, $Y_4 = Y_{imim}$ and $Y_5 = Y_{mnmn}$.
Then these are explicitly written as follows.
\begin{alignat}{3}
  Y_1 &= X_{0i0i} - \frac{1}{9} ( - X_{ii} + X_{00} ) - \frac{1}{90} X \notag
  \\
  &= \frac{64}{5} \big[ 2 (40 \!-\! 11d) W_1^3 - 12 (10\!-\!d) W_2^3 + 2 (d\!-\!1)(d^2\!-\!24d\!+\!50) W_3^3 \notag
  \\
  &\quad
  - 4 (7d\!-\!10)(10\!-\!d) W_4^3 + 2 (10\!-\!d)(9\!-\!d)(6\!-\!d) W_5^3 - 10 (10\!-\!d) W_1 W_2^2 \notag
  \\
  &\quad
  - 2 (7d\!-\!60)(d\!-\!1) W_1^2 W_3 + 5 (19d \!-\! 48)(d\!-\!1) W_1 W_3^2 + 10 (11d\!-\!12)(10\!-\!d) W_1 W_4^2 \notag
  \\
  &\quad
  + 15 (10\!-\!d)(9\!-\!d) W_1 W_5^2 + 40 (10\!-\!d) W_2^2 W_4 + 20 (d\!-\!3)(10\!-\!d) W_2 W_4^2 
  \\
  &\quad
  - 14 (10\!-\!d)(9\!-\!d) W_2^2 W_5 + 20 (10\!-\!d)(9\!-\!d) W_2 W_5^2 - 6(d\!-\!1)(10\!-\!d)^2 W_3 W_4^2 \notag
  \\
  &\quad
  + 2 (3d\!-\!10)(10\!-\!d)(9\!-\!d) W_4^2 W_5 - 20 (10\!-\!d)(9\!-\!d) W_4 W_5^2 \notag
  \\
  &\quad
  - 4 (7d\!-\!20)(10\!-\!d) W_1 W_2 W_4 + 60 (d\!-\!1)(10\!-\!d) W_2 W_3 W_4 
  + 60 (10\!-\!d)(9\!-\!d) W_2 W_4 W_5 \big]. \notag
\end{alignat}
\vspace{-1cm}
\begin{alignat}{3}
  Y_2 &= X_{0m0m} - \frac{1}{9} ( - X_{mm} + X_{00} ) - \frac{1}{90} X \notag
  \\
  &= \frac{64}{5} \big[ - 12 d W_1^3 + 2(11d\!-\!70) W_2^3 + 2 d(d\!-\!1)(d\!-\!4) W_3^3 - 4 d(60\!-\!7d) W_4^3 \notag
  \\
  &\quad
  + 2 (d^2\!+\!4d\!-\!90)(9\!-\!d) W_5^3 - 10 d W_1^2 W_2 - 14 d(d\!-\!1) W_1^2 W_3 + 20 d(d\!-\!1) W_1 W_3^2 \notag
  \\
  &\quad
  + 40 d W_1^2 W_4 + 20 d(7\!-\!d) W_1 W_4^2 + 15 d(d\!-\!1) W_2 W_3^2 + 10 d(98\!-\!11d) W_2 W_4^2 \notag
  \\
  &\quad
  + 2 (7d\!-\!10)(9\!-\!d) W_2^2 W_5 + 5 (142\!-\!19d)(9\!-\!d) W_2 W_5^2 - 20 d(d\!-\!1) W_3^2 W_4 
  \\
  &\quad
  + 2 d(d\!-\!1)(20\!-\!3d) W_3 W_4^2 - 6 d^2(9\!-\!d) W_4^2 W_5 + 4d(7d\!-\!50) W_1 W_2 W_4 \notag
  \\
  &\quad
  + 60 d(d\!-\!1) W_1 W_3 W_4 + 60 d(9\!-\!d) W_1 W_4 W_5 \big]. \notag
\end{alignat}
\vspace{-1cm}
\begin{alignat}{3}
  Y_3 &= X_{ijij} - \frac{1}{9} ( X_{jj} + X_{ii}) + \frac{1}{90} X \notag
  \\
  &= \frac{64}{5} \big[ - 8 (6d\!-\!25) W_1^3 - 8 (10\!-\!d) W_2^3 - (d\!-\!4)(2d^2 \!-\! 157d \!+\! 335) W_3^3 \notag
  \\
  &\quad
  + 8 (6d\!-\!25)(10\!-\!d) W_4^3 - 2 (10\!-\!d)(9\!-\!d)(6\!-\!d) W_5^3 - 6(d^2 \!-\! 46d \!+\! 100) W_1^2 W_3 \notag
  \\
  &\quad
  - 40 (10\!-\!d) W_1 W_2^2 + 30 (10\!-\!d) W_2^2 W_3 + 40 (10\!-\!d) W_2^2 W_4 - 6 (10\!-\!d)(9\!-\!d) W_2^2 W_5 \notag
  \\
  &\quad
  - 6 (d^2\!-\!46d\!+\!100)(10\!-\!d) W_3 W_4^2 + 15 (10\!-\!d)(9\!-\!d) W_3 W_5^2 + 40 (10\!-\!d)(9\!-\!d) W_4 W_5^2 \notag
  \\
  &\quad
  + 2 (50\!-\!3d)(10\!-\!d)(9\!-\!d) W_4^2 W_5 + 4(50\!-\!3d) (10\!-\!d) W_1 W_2 W_4 \big]. 
\end{alignat}
\vspace{-1cm}
\begin{alignat}{3}
  Y_4 &= X_{imim} - \frac{1}{9} ( X_{mm} + X_{ii}) + \frac{1}{90} X \notag
  \\
  &= \frac{64}{5} \big[ - 4 (7d\!-\!10) W_1^3 - 4(60\!-\!7d) W_2^3 - 2 (d\!-\!1)(d^2\!-\!24d\!+\!50) W_3^3 \notag
  \\
  &\quad
  - 2(69d^2 \!-\! 690d \!+\! 680) W_4^3 - 2 (d^2\!+\!4d\!-\!90)(9\!-\!d) W_5^3 - 20 (d\!-\!3) W_1^2 W_2 - 20 (7\!-\!d) W_1 W_2^2 \notag
  \\
  &\quad
  + 6 (d\!-\!1)(10\!-\!d) W_1^2 W_3 + 10 (11d \!-\! 12) W_1^2 W_4 + 10 (98\!-\!11d) W_2^2 W_4 + 6 d(9\!-\!d) W_2^2 W_5 \notag
  \\
  &\quad
  + 5 (19d\!-\!48)(d\!-\!1) W_3^2 W_4 + 2 (d\!-\!1)(3d^2\!-\!50d\!+\!210) W_3 W_4^2 
  \\
  &\quad
  + 2 (3d^2\!-\!10d\!+\!10)(9\!-\!d) W_4^2 W_5 - 5 (19d\!-\!142)(9\!-\!d) W_4 W_5^2 + 60 (d\!-\!1) W_1 W_2 W_3 \notag
  \\
  &\quad
  + 4 (3d^2\!-\!30d\!+\!70) W_1 W_2 W_4 + 60 (9\!-\!d) W_1 W_2 W_5 + 60 (d\!-\!1)(9\!-\!d) W_3 W_4 W_5 \big]. \notag
\end{alignat}
\vspace{-1cm}
\begin{alignat}{3}
  Y_5 &= X_{mnmn} - \frac{1}{9} ( X_{nn} + X_{mm}) + \frac{1}{90} X \notag
  \\
  &= \frac{64}{5} \big[ - 8 d W_1^3 - 8 (35\!-\!6d) W_2^3 - 2 d(d\!-\!1)(d\!-\!4) W_3^3 + 8 d(35\!-\!6d) W_4^3 \notag
  \\
  &\quad
  - (6-d)(2d^2\!+\!117d\!-\!1035) W_5^3 - 40 d W_1^2 W_2 - 6 d(d\!-\!1) W_1^2 W_3 + 40 d W_1^2 W_4 
  \\
  &\quad
  + 30 d W_1^2 W_5 - 6 (d^2\!+\!26d\!-\!260) W_2^2 W_5 + 40 d(d\!-\!1) W_3^2 W_4 + 2 d(d\!-\!1)(3d\!+\!20) W_3 W_4^2 \notag
  \\
  &\quad
  + 15 d(d\!-\!1) W_3^2 W_5 - 6 d(d^2\!+\!26d\!-\!260) W_4^2 W_5 + 4 d(3d\!+\!20) W_1 W_2 W_4 \big]. \notag
\end{alignat}

Fifth, let us evaluate $D_c D_d Y^c{}_{ab}{}^d$.
Note that the nonzero components of the spin connection are $\omega_i{}^i{}_0 = H$ and $\omega_m{}^m{}_0 = G$,
and terms with spatial derivatives become zero.
Then $D_c D_d Y^c{}_{ab}{}^d$ is simplified as follows.
\begin{alignat}{3}
  D_c D_d Y^c{}_{ab}{}^d 
  &= \partial_c (\partial_d Y^c{}_{ab}{}^d + \omega_d{}^c{}_e Y^e{}_{ab}{}^d - \omega_d{}^e{}_a Y^c{}_{eb}{}^d 
  - \omega_d{}^e{}_b Y^c{}_{ae}{}^d +  \omega_d{}^d{}_e Y^c{}_{ab}{}^e) \notag
  \\
  &\quad\,
  + \omega_c{}^c{}_e (\partial_d Y^e{}_{ab}{}^d + \omega_d{}^e{}_f Y^f{}_{ab}{}^d - \omega_d{}^f{}_a Y^e{}_{fb}{}^d 
  - \omega_d{}^f{}_b Y^e{}_{af}{}^d + \omega_d{}^d{}_f Y^e{}_{ab}{}^f) \notag
  \\
  &\quad\,
  - \omega_c{}^e{}_a (\partial_d Y^c{}_{eb}{}^d + \omega_d{}^c{}_f Y^f{}_{eb}{}^d - \omega_d{}^f{}_e Y^c{}_{fb}{}^d 
  - \omega_d{}^f{}_b Y^c{}_{ef}{}^d + \omega_d{}^d{}_f Y^c{}_{eb}{}^f) \notag
  \\
  &\quad\,
  - \omega_c{}^e{}_b (\partial_d Y^c{}_{ae}{}^d + \omega_d{}^c{}_f Y^f{}_{ae}{}^d - \omega_d{}^f{}_a Y^c{}_{fe}{}^d 
  - \omega_d{}^f{}_e Y^c{}_{af}{}^d + \omega_d{}^d{}_f Y^c{}_{ae}{}^f) \notag
  \\
  &= (\partial_0 + \omega_c{}^c{}_0) (\partial_0 Y^0{}_{ab}{}^0 + \omega_d{}^0{}_d Y^d{}_{ab}{}^d - \omega_d{}^d{}_a Y^0{}_{db}{}^d 
  - \omega_a{}^0{}_b Y^0{}_{a0}{}^a + \omega_d{}^d{}_0 Y^0{}_{ab}{}^0) \notag
  \\
  &\quad\,
  - \omega_b{}^0{}_a \partial_0 Y^b{}_{0b}{}^0 + \omega_b{}^0{}_a \omega_d{}^d{}_0 Y^b{}_{db}{}^d 
  + \omega_b{}^0{}_a \omega_b{}^0{}_b Y^b{}_{00}{}^b - \omega_b{}^0{}_a \omega_d{}^d{}_0 Y^b{}_{0b}{}^0 
  \\
  &\quad\,
  - \omega_c{}^c{}_b \partial_0 Y^c{}_{ac}{}^0 + \omega_c{}^c{}_b \omega_d{}^d{}_a Y^c{}_{dc}{}^d 
  + \omega_c{}^c{}_b \omega_c{}^0{}_c Y^c{}_{a0}{}^c - \omega_c{}^c{}_b \omega_d{}^d{}_0 Y^c{}_{ac}{}^0. \notag
\end{alignat}
Note that $\partial_0 = \partial_t$. 
Nonzero components of $D_c D_d Y^c{}_{ab}{}^d$ are calculated as follows.
\begin{alignat}{3}
  D_c D_d Y^c{}_{00}{}^d &= 
  \omega_c{}^c{}_0 \partial_0 Y_{c0c0} + \omega_c{}^c{}_0 \omega_d{}^d{}_0 Y_{cdcd} 
  - (\omega_c{}^c{}_0)^2 Y_{0c0c} + \omega_c{}^c{}_0 \omega_d{}^d{}_0 Y_{c0c0} \notag
  \\
  &= d H \dot{Y}_1 + (10\!-\!d) G \dot{Y}_2 + d(d\!-\!1) H^2 (Y_1 + Y_3) \notag
  \\
  &\quad\,
  + (10\!-\!d)(9\!-\!d) G^2 (Y_2 + Y_5) + d(10\!-\!d) HG (Y_1 + Y_2 + 2 Y_4), \notag
  \\
  D_c D_d Y^c{}_{ii}{}^d 
  &= (\partial_0 + \omega_c{}^c{}_0) ( - \partial_0 Y_{0i0i} - \omega_d{}^0{}_d Y_{idid} + \omega_i{}^0{}_i Y_{0i0i} - \omega_d{}^d{}_0 Y_{0i0i}) \notag
  \\
  &\quad\,
  + \omega_i{}^0{}_i \partial_0 Y_{i0i0} + \omega_i{}^0{}_i \omega_d{}^d{}_0 Y_{idid}
  - (\omega_i{}^0{}_i)^2 Y_{0i0i} + \omega_i{}^0{}_i \omega_d{}^d{}_0 Y_{i0i0} \label{eq:DDYcal}
  \\
  &= \Big\{ \frac{d}{dt} + (d\!-\!1) H + (10\!-\!d) G \Big\} 
  \{ - \dot{Y}_1 - (d\!-\!1) H (Y_1 + Y_3) - (10\!-\!d) G (Y_1 + Y_4) \}, \notag
  \\
  D_c D_d Y^c{}_{mm}{}^d 
  &= (\partial_0 + \omega_c{}^c{}_0) ( - \partial_0 Y_{0m0m} - \omega_d{}^0{}_d Y_{mdmd} + \omega_m{}^0{}_m Y_{0m0m} - \omega_d{}^d{}_0 Y_{0m0m}) \notag
  \\
  &\quad\,
  + \omega_m{}^0{}_m \partial_0 Y_{m0m0} + \omega_m{}^0{}_m \omega_d{}^d{}_0 Y_{mdmd}
  - (\omega_m{}^0{}_m)^2 Y_{0m0m} + \omega_m{}^0{}_m \omega_d{}^d{}_0 Y_{m0m0} \notag
  \\
  &= \Big\{ \frac{d}{dt} + d H + (9\!-\!d) G \Big\} \{ - \dot{Y}_2 - d H (Y_2 + Y_4) - (9\!-\!d) G (Y_2 + Y_5) \}. \notag
\end{alignat}

Sixth, components of $R_{cdea} Y^{cde}{}_b = 2 R_{eaea} Y^{eae}{}_b$ are evaluated as follows.
\begin{alignat}{3}
  R_{cde0} Y^{cde}{}_0 &= - 2 R_{e0e0} Y_{e0e0} 
  = 2 d (\dot{H} + H^2) Y_1 + 2 (10\!-\!d) (\dot{G} + G^2) Y_2, \notag
  \\
  R_{cdei} Y^{cde}{}_i &= 2 R_{eiei} Y_{eiei} 
  = - 2 (\dot{H} + H^2) Y_1 + 2 (d\!-\!1) H^2 Y_3 + 2 (10\!-\!d) HG Y_4, \label{eq:RYcal}
  \\
  R_{cdem} Y^{cde}{}_m &= 2 R_{emem} Y_{emem} 
  = - 2 (\dot{G} + G^2) Y_2 + 2 d HG Y_4 + 2 (9\!-\!d) G^2 Y_5. \notag
\end{alignat}

Finally, combining the eqs.~(\ref{eq:Zcal}), (\ref{eq:DDYcal}) and (\ref{eq:RYcal}),
we obtain $\Gamma$ dependent part in the eq.~(\ref{eq:MEOMgen}).


\end{document}